# Spin-orbiton and quantum criticality in FeSc$_2$S$_4$

L. Mittelstädt,[1] M. Schmidt,[1] Zhe Wang,[1] F. Mayr,[1] V. Tsurkan,[1,2] P. Lunkenheimer,[1,*]
D. Ish,[4] L. Balents,[3] J. Deisenhofer,[1] and A. Loidl[1]

[1]*Experimental Physics V, Center for Electronic Correlations and Magnetism, University of Augsburg, 86135 Augsburg, Germany*
[2]*Institute of Applied Physics, Academy of Sciences of Moldova, MD-2028 Chisinau, Republic of Moldova*
[3]*Kavli Institute for Theoretical Physics, University of California, Santa Barbara, California, 93106-4030, USA*
[4]*Physics Department, University of California, Santa Barbara, California, 93106-4030, USA*

In FeSc$_2$S$_4$ spin-orbital exchange competes with strong spin-orbit coupling, suppressing long-range spin and orbital order and, hence, this material represents one of the rare examples of a spin-orbital liquid ground state. Moreover, it is close to a quantum-critical point separating the ordered and disordered regimes. Using THz and FIR spectroscopy we study low-lying excitations in FeSc$_2$S$_4$ and provide clear evidence for a *spin-orbiton*, an excitation of strongly entangled spins and orbitals. It becomes particularly well pronounced upon cooling, when advancing deep into the quantum-critical regime. Moreover, indications of an underlying structureless excitation continuum are found, a possible signature of quantum criticality.

PACS numbers: 71.70.Ej, 78.30.Hv, 75.40.Gb

## I. INTRODUCTION

In frustrated magnets, conventional long-range spin order is suppressed via competing interactions or geometrical frustration. Frustration leads to macroscopically degenerate ground states, susceptible to the emergence of exotic magnetic states, like, e.g., spin liquids, in which the strongly coupled spin system fluctuates down to temperatures of absolute zero.[1] The cubic spinel structure with stoichiometry AB$_2$X$_4$, one of the most frequently encountered structures among transition metal oxides, provides prototypical examples for frustrated spin systems: The octahedrally coordinated *B* sites constitute a pyrochlore lattice, in three dimensions one of the strongest contenders of frustration with spin-liquid or spin-ice like ground states. The tetrahedrally coordinated *A* sites form a diamond lattice, another prominent example for frustration: It consists of two interpenetrating fcc lattices and frustration occurs with respect to the ratio of inter- and intra-lattice magnetic exchange. The spinel MnSc$_2$S$_4$, with manganese at the *A* site and nonmagnetic Sc at the *B* site, is an illuminating example where a spiral spin-liquid state evolves at low temperatures.[2,3] Spin-liquid states have also been identified in a series of aluminum based A-site spinels.[4,5,6]

However, not only the spins but also the orbital degrees of freedom can evade long-range order. In FeSc$_2$S$_4$, Fe$^{2+}$ with a d$^6$ electronic configuration is tetrahedrally coordinated by S$^{2-}$, and consequently reveals a twofold orbital degeneracy. A Jahn-Teller transition leading to long-range orbital order is expected at low temperatures. However, FeSc$_2$S$_4$ neither shows orbital nor magnetic order down to 50 mK, despite the natural energy scales set by the magnetic exchange of 45 K (Ref. 7) and by the Jahn-Teller energy of 10 K of tetrahedrally coordinated Fe$^{2+}$, as detected in the isostructural chromium spinel FeCr$_2$S$_4$ (Ref. 8). The fact that in these systems, either by frustration or by disorder, orbital order can indeed be easily suppressed, has been documented in Ref. 9, where a low-temperature orbital glass state has been identified. In FeSc$_2$S$_4$ orbital and spin order are suppressed almost down to zero temperature and hence, it represents one of the rare examples of a spin-orbital liquid (SOL).[7,10,11,12,13] An anomalously small excitation gap in this material was identified by NMR[14] and neutron scattering.[15]

Interestingly, based on theoretical considerations, it has been suggested that the SOL state in FeSc$_2$S$_4$ does not result from frustration but from a competition between on-site spin-orbit coupling (SOC) and spin-orbital exchange.[1,10,11] Generally, the coupling of electronic spin and orbital momentum is of prime importance in atomic physics, plays a significant role in condensed matter,[16] and even can be used in spin-orbitronics: the simultaneous manipulation of both these electronic degrees of freedom.[17,18] While spin and orbital exchange favors order, strong SOC can result in a SOL state with high entanglement of the spin and orbital subsystems.[1,10,11] According to this theory, FeSc$_2$S$_4$ is close to a quantum-critical (QC) point between the SOL state and a magnetically and orbitally ordered phase. This is indicated in Fig. 1 showing a schematic quantum critical phase diagram, where the control parameter $x$ represents the ratio of magnetic exchange $J$ to the effective spin-orbit interaction $\lambda$ (Refs. 1,11). A quantum-critical region separates a disordered, so-called spin-orbital singlet state and a state with antiferromagnetic and orbital order. The vertical dashed line indicates the position of FeSc$_2$S$_4$, for which the QC regime is estimated to arise between about 2 and 45 K (Ref. 11). Quantum criticality provides a new organizing principle in condensed matter physics and can control sizable regions of phase diagrams with far-reaching consequences, including a plethora of exotic phases.[19] Hence, FeSc$_2$S$_4$ not only belongs to the very few examples of a spin-orbital liquid which is induced by strong spin-orbital entanglement, but the strength of its SOC brings it also close to a quantum-critical point (Fig. 1). Thus, it seems worthwhile to study the low-lying excitation spectrum in this system.

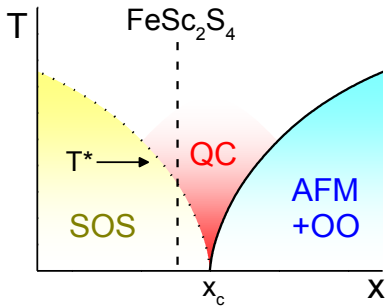

FIG. 1 (color online). Schematic x,T phase diagram based on theoretical considerations[1,11] (see text for details). The control parameter $x$ measures the ratio of the magnetic exchange to the effective spin-orbit coupling. SOS, QC, AFM, and OO denote spin-orbital singlet, quantum critical, antiferromagnetic, and orbitally ordered states, respectively. The vertical dashed line indicates the position of $FeSc_2S_4$.

In order to understand the excitation spectrum of the coupled spin and orbital degrees of freedom of $FeSc_2S_4$, it is useful first to consider the crystal-field splitting including spin-orbit coupling of $Fe^{2+}$ impurities in tetrahedral symmetry. The crystal-field splitting has been calculated in detail by Low and Wegener[20] and the splitting of the crystal-field ground state $^5E$ into five levels equally separated by $\lambda$, with $\lambda$ being the effective spin-orbit interaction $\lambda = 6\lambda_0^2/\Delta$, has been mainly studied by optical spectroscopy and elucidated by various authors.[21,22,23,24,25] Here $\lambda_0$ is the atomic SOC and $\Delta$ the crystal-field splitting. This splitting of the ground state by second order spin-orbit interaction of $Fe^{2+}$ ($3d^6$) in tetrahedral symmetry has been determined to be of order $\lambda \approx 15 - 20$ cm$^{-1}$ (Refs. 21,23,24,25), with the ground state being a singlet and the first excited state being a triplet state. Assuming formal selection rules, the states within the $^5E$ manifold are connected via a magnetic-dipole transition only. However, due to the admixture of the wave functions of a higher crystal-field level, electric-dipole transitions become possible, too. In the single-ion cases studied so far, a magnetic dipole transition has been observed at $\lambda$, while an electric dipole transition was identified at $3\lambda$.[21]

No similar optical experiments on SOC have been performed on concentrated systems, i.e. spin and orbital moments on regular lattice sites. Here, due to strong exchange interactions strong dispersion effects of the entangled spin-orbital degrees of freedom are expected and indeed have been observed in $FeSc_2S_4$ by neutron scattering:[15] Strongly dispersing modes reveal a small gap at the hypothetical AFM zone boundary of order 0.2 meV ≈ 1.6 cm$^{-1}$, by a factor of 10 smaller than the single-ion value of an excitation within the $^5E$ manifold. The wave-vector dependence of the low-lying singlet-triplet excitation has been calculated in Ref. 10. It is a strongly propagating mode, being almost soft at the antiferromagnetic zone boundary and strongly enhanced at the zone center. The zone center value of $1.9\lambda$ should be detectable by THz spectroscopy.

In the present work, we provide THz- and far-infrared (FIR) spectroscopy results for $FeSc_2S_4$ to search for such theoretically predicted generic excitations of SOLs,[10,11] which we term spin-orbitons, and for signatures of the QC state. Pure orbital excitations, so-called orbitons, have been observed earlier in orbitally ordered $LaMnO_3$ (Ref. 26) and vanadium oxides.[27] These crystal-field derived excitations are located in the infrared region, while in the present work we focus on the ground-state splitting due to spin-orbit coupling at much lower energies, in the THz regime.

## II. EXPERIMENTAL DETAILS

Samples of $FeSc_2S_4$ were prepared by sintering stoichiometric mixtures of the high-purity elements Fe (4N), Sc (3N) and S (5N) in evacuated sealed silica ampoules at 1000°C. After a sintering time of one week, the samples were powdered, homogenized, pressed into pellets, and annealed again at 1000°C, a synthesis procedure which was repeated several times to reach full reaction. The samples were characterized by X-ray diffraction, magnetic susceptibility, and heat-capacity experiments, resulting in structural, magnetic, and thermodynamic properties as described in detail by Fritsch et al.[7]

Time-domain THz transmission experiments were carried out between liquid helium and room temperature using a TPS Spectra 3000 spectrometer (TeraView Ltd.). Reflectivity experiments in the FIR range were performed in the same temperature range using a Bruker Fourier-transform spectrometer IFS 113v equipped with a He-flow cryostat. With the set of mirrors and detectors used for these experiments, we were able to cover a frequency range from 60 to 700 cm$^{-1}$. For the reflectivity experiments, the ceramics were pressed with a maximum pressure of 1 GPa with the surfaces polished to optical quality. Nevertheless, neither surface nor density of the samples was ideal and of perfect optical quality and we were not able to receive reliable results below 80 cm$^{-1}$. In THz spectroscopy, real and imaginary part of the dielectric permittivity can be directly derived from transmission and phase shift and we determined absolute values of the complex permittivity. We had to correct all FIR reflectivity spectra by a factor of 1.3 to bring in line the permittivity from the FIR and THz results. This correction was necessary to account for low densities and non-ideal surfaces of the ceramic samples. All reflectivity spectra were fitted using the RefFIT fit routine, including the option for Fano-type resonance line shapes.[28] More details can be found in the Supplemental Material.[29]

## III. EXPERIMENTAL RESULTS AND DISCUSSION

Figure 2 shows the dielectric loss of $FeSc_2S_4$ for wave numbers between 15 and 65 cm$^{-1}$ and temperatures between 6 and 50 K in a three-dimensional plot. This figure impressively shows how, on decreasing temperatures a low-lying excitation located roughly at 30 cm$^{-1}$ evolves from a structureless continuum with almost linear energy dependence. At 5 K it is located at 35 cm$^{-1}$, significantly above the single-ion value $\lambda$ and indicating strong exchange interactions. Below about 20 K, this excitation becomes exceedingly sharper, while for higher temperatures it is almost overdamped. The oscillator strength of this excitation is rather weak and, using hand-waving arguments, one would assume this mode to result from a magnetic dipole transition, as expected for a transition within the $^5E$ manifold.

Figure 3(a) shows the frequency dependence of the dielectric loss as documented in Fig. 2 for selected temperatures between 4 and 70 K. At low temperatures a sharp excitation close to 35 cm$^{-1}$ evolves from a continuously increasing background. We ascribe this mode to a spin-orbiton, an excitation of strongly entangled spin and orbital degrees of freedom.[10,11] We would like to recall that this spin-orbiton is almost soft at the AFM zone boundary[15] and in the single-ion case this excitation is expected between 15 and 20 cm$^{-1}$. For temperatures ≳ 20 K, this excitation becomes rather smeared out. At lower temperatures its eigenfrequency significantly shifts towards higher wave numbers and the damping strongly decreases. At frequencies > 50 cm$^{-1}$ and the lowest temperatures, the loss increases more strongly, possibly indicating the appearance of a second excitation. In the single-ion case an electric dipole excitation is expected close to 3 $\lambda$ ~ 45 - 60 cm$^{-1}$.

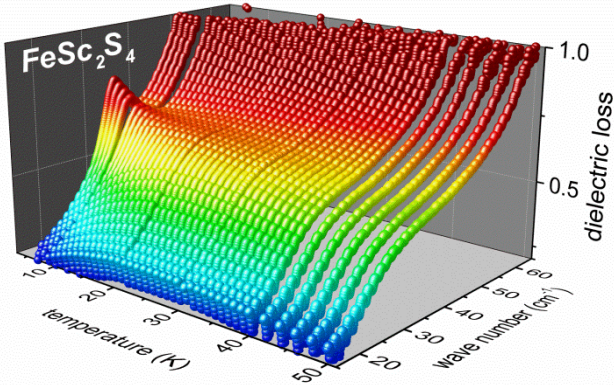

FIG. 2 (color). Three dimensional plot of the dielectric loss of FeSc$_2$S$_4$ vs. wave number and temperature. The loss is color coded to indicate equal-loss contours above the temperature wave-number plane.

For a more quantitative analysis, we fitted the data to the sum of a Lorentzian line shape and an increasing background, the latter estimated by a spline interpolation between the regions just outside the excitation peak. The fit curves [lines in Fig. 3(a)] are in good agreement with experiment. The resulting parameters of the Lorentzian-shaped peaks are provided in Figs. 3(b) - (d), all showing significant changes of temperature characteristics below $T^*$ ~ 17 K (vertical line), where a significant increase in eigenfrequency is accompanied by a strong decrease in damping. $T^*$ is located within the QC regime as indicated by the arrow in the schematic phase diagram, shown in Fig. 1. At this temperature the spin-orbiton evolves as a well-defined coherent excitation. In the single ion case the eigenfrequency should remain constant and the oscillator strength is expected to continuously decrease on increasing temperature, due to the thermal population of higher orbitals within the ground state manifold. The non-monotonous temperature dependence of all parameters of the Lorentzian line signals the importance of spin-orbital entanglement and indicates a characteristic temperature $T^*$.

The found THz excitation can directly be compared to model calculations of the SOL in FeSc$_2$S$_4$. According to Refs. 10 and 11, SOC splits the local spin and orbital degeneracy to form entangled spin-orbital states, with a spin-orbital singlet ground state. As outlined earlier, these states in turn are renormalized by exchange, such that the triplet magnetic excitations acquire strong dispersion, being soft close to the zone boundary, corresponding to a wave vector characteristic for hypothetical antiferromagnetic order and with strongly enhanced eigenfrequencies at the zone center. In optical spectroscopy, being sensitive to excitations at the zone center, a magnetic dipole mode is expected at $\lambda$ and an electric dipole excitation at 3 $\lambda$. According to Ref. 10, the exchange-induced dispersion renormalizes the triplet mode at $\lambda$ to a maximum at the zone center of approximately 1.9 $\lambda$. While the renormalization of the electric dipole excitation has not been studied theoretically, some upward shift is also expected at zero wave vectors and an electric dipole-active mode should exist above 45 – 60 cm$^{-1}$.

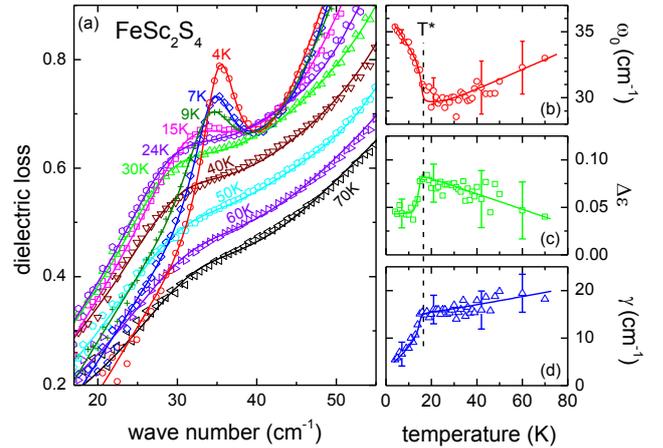

FIG. 3 (color online). (a) Dielectric loss of FeSc$_2$S$_4$ at low wave numbers for selected temperatures between 4 and 70 K. The solid lines represent fits using Lorentzian line shapes including a temperature-dependent background (see text). Right column: Temperature dependence of eigenfrequency (b), of the dielectric strength (c), and of the damping constant (d) of the spin-orbiton. The solid lines in (b), (c), and (d) serve as guides to the eye. The dashed vertical line indicates a temperature $T^* \approx 17$ K, below which a change of the characteristics of the excitation is observed.

Comparing the evolution of a well-defined excitation close to 35 cm$^{-1}$ - shown in Fig. 3(a) - to the theoretical estimate of ≈ 30 - 40 cm$^{-1}$ gives confidence to its interpretation in terms of a renormalized magnetic dipole excitation between the singlet ground state and the triplet excited state. Interestingly, the eigenfrequeny of this excitation significantly increases and its damping decreases below $T^* \approx 17$ K [Figs. 3(b) - (d)], i.e. within the QC region (Fig. 1), which was estimated to emerge between about 2 and 45 K (Ref. 11). This change of temperature characteristics is consistent with the formation of the SOL state and most likely signals the gradual loss of decay channels as one advances deeper into the QC regime. The fact that this excitation corresponds to the singlet-triplet excitation in the atomic limit has also been verified by THz experiments under external magnetic fields showing a clear splitting of the mode.[30]

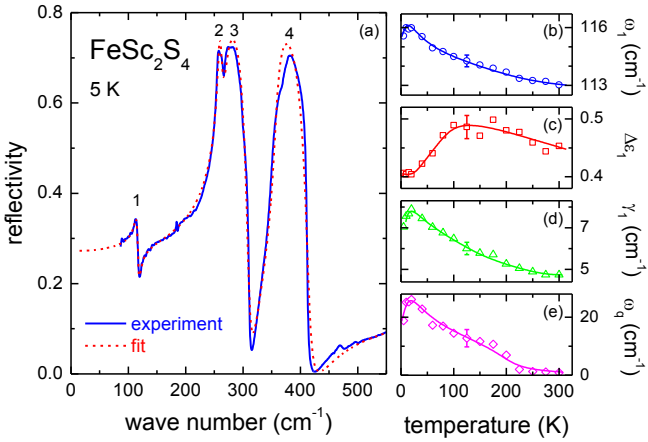

FIG. 4 (color online). (a) FIR measurements between 80 and 600 cm$^{-1}$ are compared with a fit utilizing four Lorentz oscillators. The four phonon modes are indexed from 1 to 4. The right frames show the temperature dependence of eigenfrequency $\omega_0$ (b), dielectric strength $\Delta\varepsilon$ (c), damping $\gamma$ (d), and the Fano factor $\omega_q$ (e) of phonon 1 close to 115 cm$^{-1}$. In (b) to (e), typical error bars are indicated. The solid lines are drawn to guide the eyes.

To probe a possible coupling of phonon modes to the spin-orbital excitations, to search for a possible second - electric-dipole active - spin-orbital excitation proposed theoretically, and to clarify the origin of the linearly increasing background in the THz regime, we measured the reflectivity of FeSc$_2$S$_4$ in the FIR region between room and liquid-helium temperatures. A representative result of the reflectivity as observed at 5 K for wave numbers between 80 and 550 cm$^{-1}$ is shown in Fig. 4(a). Room-temperature FIR spectra of FeSc$_2$S$_4$ have been published previously with five so far unexplained IR-active phonon excitations.[31] Our results as documented in Fig. 4(a) reveal four prominent reflectivity bands, characteristic for normal spinel compounds.[32] The narrow spike close to 180 cm$^{-1}$ is an experimental artifact and the weak mode close to 470 cm$^{-1}$, earlier interpreted as phonon mode,[31] could be the reflectance due an electronic quadrupolar excitation,[33] but is not further analyzed in the course of this work.

The experimental reflectivity is fitted assuming four Lorentz oscillators characterizing the four optical phonons expected for normal spinel compounds. Phonon 1, close to 115 cm$^{-1}$, can only be satisfactorily described assuming an asymmetric Fano-type line shape as resulting from interference of resonant scattering with a broad excitation continuum[34] (see also Supplemental Material[29]). Hence, as described earlier, all reflectivity spectra were fitted using the RefFIT fit routine,[28] including the option for Fano-type resonance line shapes.

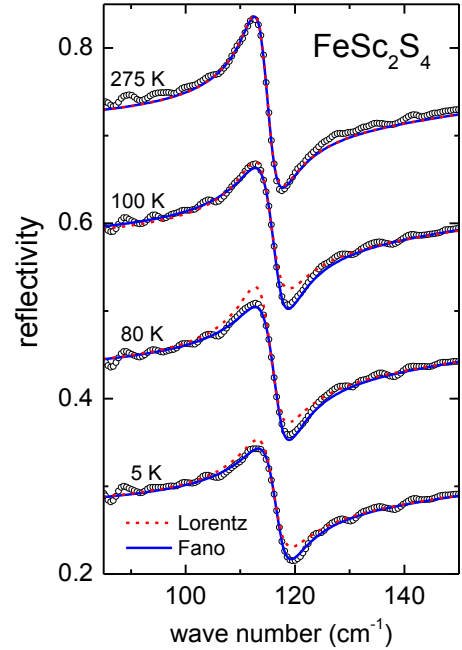

FIG. 5 (color online). Temperature dependence of the reflectivity in FeSc$_2$S$_4$ in the frequency regime of the lowest phonon mode. For clarity reasons, the reflectivity curves for 80, 100, and 275 K are shifted upwards. The reflectivity of phonon mode 1 of FeSc$_2$S$_4$ was fitted with a pure Lorentz oscillator model (dashed red lines) or with a Fano resonance model, eq. (1) (solid blue lines).

Adapting this Fano-type phenomenon, the equation for the complex dielectric constant as derived from a Lorentz oscillator has to be modified, including a Fano parameter $\omega_q$:

$$\epsilon(\omega) = \frac{\omega_p^2}{\omega_0^2-\omega^2-i\gamma\omega}\left(1+i\frac{\omega_q}{\omega}\right)^2 + \left(\frac{\omega_p\omega_q}{\omega_0\omega}\right)^2 \quad (1)$$

Here $\omega_o$ is the phonon eigenfrequency, $\omega_p$ is the ionic plasma frequency, $\gamma$ the damping constant of the phonon modes, describing the inverse life time of phonon excitations, and $\omega_q$ is the Fano factor, describing the asymmetry of the phonon line shape. From the ionic plasma frequency, the dielectric strength $\Delta\varepsilon = (\omega_p/\omega_0)^2$ can be calculated. It is clear from eq. (1) that for $\omega_q = 0$ the Lorentzian line shape is recovered. We found that for the lowest phonon mode (phonon 1) below 100 K, reasonable fits can only be obtained including a temperature-dependent Fano factor. In contrast, the fits close to room temperature are hardly improved and there satisfactory fits can also be obtained utilizing pure Lorentzian line shapes. This is documented in Fig. 5, which shows representative results of the reflectivity for the phonon mode close to 115 cm$^{-1}$ and demonstrates the importance of the Fano factor $\omega_q$ for a series of temperatures.

A corresponding analysis as in Fig. 4(a) was performed for further temperatures up to 300 K, yielding temperature-dependent eigenfrequencies, oscillator strengths, damping constants, and (for phonon 1) the Fano factor. Phonon modes 2-4 exhibit temperature-dependent parameters, typical for canonical anharmonic modes (see Supplemental Material[29]). However, for the phonon mode at 115 cm$^{-1}$ the eigenfrequency [Fig. 4(b)] shows no sign of saturation at low temperatures and the damping constant [Fig. 4(d)] significantly increases on decreasing temperature, both findings contrary to canonical anharmonic behavior. These unconventional temperature dependencies signal significant coupling of this phonon mode to other excitations, which in FeSc$_2$S$_4$ certainly are due to the spin and orbital degrees of freedom.

Most interestingly, the oscillator strength of this mode [Fig. 4(c)] exhibits a slight increase from room temperature down to 100 K, but then decreases by almost 20 % upon further cooling. It seems reasonable that some optical weight from the low-frequency phonons is transferred to the spin-orbital excitations. The situation can be compared to electromagnon excitations of multiferroics, where significant weight from optical phonons is transferred to pure spin waves.[35] Indeed, at least some fraction of the optical weight reappears in the spin-orbiton: A comparison of Figs. 3(c) and 4(c) reveals that the phonon decreases in optical strength by 0.1, while the spin-orbiton, appearing at a three times lower frequency, has a dielectric strength of order 0.05 only. Obviously, a large fraction of the optical weight is also transferred into the continuum or in a possible second spin-orbital excitation. The Fano factor [Fig. 4(e)] approaches zero at high temperatures but becomes large at low temperatures. Notably, the temperature dependences of all parameters reveal anomalies close to 15-20 K. They correspond to the variations in the spin-orbital excitation below about 17 K, documented in Fig. 3.

Further significant information can be obtained from Fig. 6 showing the combined THz and FIR loss spectra at 5 and 80 K: i) At 80 K, an almost linear increase at frequencies below 60 cm$^{-1}$ forms a continuum below the spin-orbiton excitation peak. In contrast, at 5 K a stronger superlinear background is found. ii) Phonon mode 1 has a strongly asymmetric Fano line shape, in agreement with fits of the reflectivity. It can be assumed to arise from interference of the resonant lattice vibration with the continuum of the SOL discussed in the following paragraph. iii) Optical weight of the phonon mode at 115 cm$^{-1}$ becomes suppressed at low temperatures [cf. Fig. 4(c)] and obviously is partly transferred to the spin-orbital excitation. iv) At 5 K and frequencies beyond the first spin-orbiton peak, excess intensity emerges. Interestingly, just in this spectral region (at frequencies higher than $3\lambda \approx 45 - 60$ cm$^{-1}$) the presence of a second spin-orbiton excitation is theoretically predicted as discussed above. However, at present it cannot be excluded that this excess intensity arises from the mentioned superlinear background and/or from the Fano-type low-frequency wing of the phonon mode. Only future experimental work, closing the frequency gap between about 60 and 85 cm$^{-1}$, can provide a definite proof of a possible second spin-orbital excitation in this frequency regime.

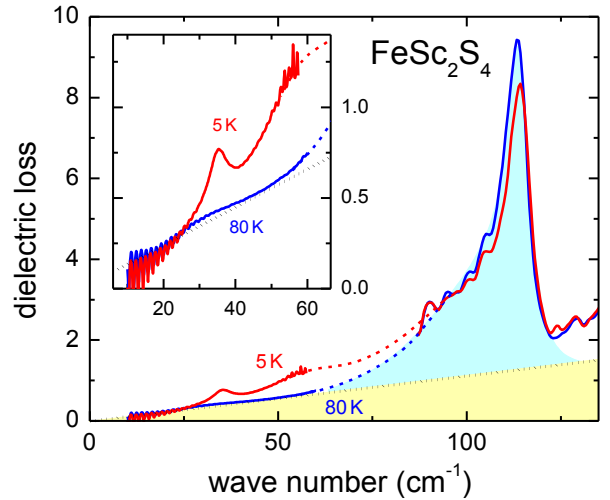

FIG. 6 (color). Low-frequency dielectric loss of FeSc$_2$S$_4$. Combined THz and FIR results of the dielectric loss are shown for wave numbers between 10 and 135 cm$^{-1}$, covering spin-orbital excitations and the lowest phonon mode. The results at 5 K are compared to measurements at 80 K. The high-temperature spectrum is composed of a linear excitation continuum (yellow) and a Fano-type phonon mode (blue). As revealed by the inset, showing a magnified view of the low-frequency results, a heavily damped spin-orbital mode is superimposed to the continuum. At 5 K, excess intensity is observed, consistent with a spin-orbiton emerging from a superlinear background. The dashed lines are guides for the eyes.

The found presence of continuum weight in the optical conductivity is actually an expected signature of quantum criticality, which induces scale invariant power-law behavior of many physical properties. In FeSc$_2$S$_4$ at high temperatures, the dielectric loss clearly exhibits an underlying linear increase in frequency. According to theory, in the QC regime the SOL is described by a multi-component $\varphi^4$ theory, where $\varphi_a$ are the components of the antiferromagnetic order parameter.[10,11] To determine the contribution of QC modes to the dielectric constant, we require the relation of the electric polarization $P_a$ to the order parameter. With some reasonable assumptions, a symmetry analysis implies $P_a \propto C_{abcd}\,\varphi_b\,\partial_c\,\varphi_d$, where $C$ is a non-zero tensor. Then the contribution to the dielectric constant is $\Delta\varepsilon \propto \langle P(\omega)\,P(-\omega)\rangle$. This correlation function can be calculated following standard methods,[36] which gives $\varepsilon''(\omega) \propto \omega^2 \coth(\omega/4k_BT)$, where $k_B$ is Boltzmann's constant, for frequencies well above the gap. Physically, the continuum arises due to the contribution of *pairs* of triplet excitations. Note that the above form is only strictly valid when $\omega$ is small compared to the magnetic bandwidth, but it nevertheless qualitatively explains the linear dielectric background found at high temperatures elegantly. Moreover, for low temperatures the coth factor becomes constant leading to a superlinear increase, in accord with the experimental findings (Fig. 6). However, it should be noted that the experimentally observed temperature dependence of the continuum is not well reproduced by theory (when assuming a temperature-independent prefactor) as the coth term decreases with decreasing temperature.

## IV. SUMMARY

In summary, in this work we have presented a combined THz and FIR study of $FeSc_2S_4$, a material belonging to the very few examples with a spin-orbital liquid groundstate, which, in addition, is characterized by a spin-orbit coupling that brings it very close to a quantum-critical point.[10]

In the course of this work we have obtained a number of remarkable results and features: We have clearly identified a low-lying spin-orbital excitation in $FeSc_2S_4$, termed spin-orbiton, i.e. propagating waves of entangled spin and orbital degrees of freedom. At low temperatures, this excitation is located close to 35 cm$^{-1}$, by a factor of 2 enhanced compared to a single-ion spin-orbit excitation, in good agreement with theoretical predictions.[10] The spin-orbiton changes character below $T^* \approx 17$ K, deep within the QC regime, becoming a well-defined long-lived excitation, which may indicate an unexpected fine structure of decay channels. Below this temperature, a highly damped mode develops into a narrow coherent excitation. From a detailed study of the temperature dependence of the phonon modes we derive a rather unconventional temperature dependence of eigenfrequency, oscillator strength, and damping of the lowest phonon mode, which signals strong coupling of lattice vibrations with the spin-orbital excitations. There is clear evidence for the transfer of optical weight from the lowest phonon mode to the spin-orbital excitations. Hence, the spin-orbiton close to 35 cm$^{-1}$ can be characterized as magnetic dipole excitation, but obviously gains some dipolar weight via coupling to the lowest phonon mode.

In addition, Fig. 6 of the present work provides some experimental evidence for a further spin-orbital excitation close to 60 cm$^{-1}$. In the single-ion case an electric dipolar excitation is expected close to $3\lambda \approx 45 - 60$ cm$^{-1}$. Moreover, we detected a temperature-dependent underlying background contribution to the dielectric loss, which seems to indicate a spin-orbital excitation continuum, a characteristic signature of quantum criticality.[10,11] The characteristic Fano line shape of the lowest phonon mode obviously stems from the interference of resonant phonon scattering with this spin-orbital continuum. Explaining these finding represents a challenge for future theoretical and experimental exploration.

## ACKNOWLEDGMENTS

We thank N. P. Armitage for helpful comments. This work was supported by the Deutsche Forschungsgemeinschaft via the Transregional Collaborative Research Center TRR 80 (Augsburg/Munich/Stuttgart). D.I. and L.B. were supported by the U.S. D.O.E. grant DE-FG02-08ER46524.

# Supplemental Material

# Spin-orbiton and quantum criticality in FeSc$_2$S$_4$

L. Mittelstädt,[1] M. Schmidt,[1] Zhe Wang,[1] F. Mayr,[1] V. Tsurkan,[1,2] P. Lunkenheimer,[1,*]
D. Ish,[4] L. Balents,[3] J. Deisenhofer,[1] and A. Loidl[1]

[1]*Experimental Physics V, Center for Electronic Correlations and Magnetism, University of Augsburg, 86135 Augsburg, Germany*
[2]*Institute of Applied Physics, Academy of Sciences of Moldova, MD-2028 Chisinau, Republic of Moldova*
[3]*Kavli Institute for Theoretical Physics, University of California, Santa Barbara, California, 93106-4030, USA*
[4]*Physics Department, University of California, Santa Barbara, California, 93106-4030, USA*

*e-mail: peter.lunkenheimer@physik.uni-augsburg.de

This supplemental material to the manuscript *Spin-orbiton and quantum criticality in FeSc$_2$S$_4$* provides more detailed information on the results from THz spectroscopy and on the analysis of FIR reflectivity spectra using Lorentzian oscillators including Fano-like line shapes. In addition, more detailed experimental details are provided.

## 1. THz spectroscopy

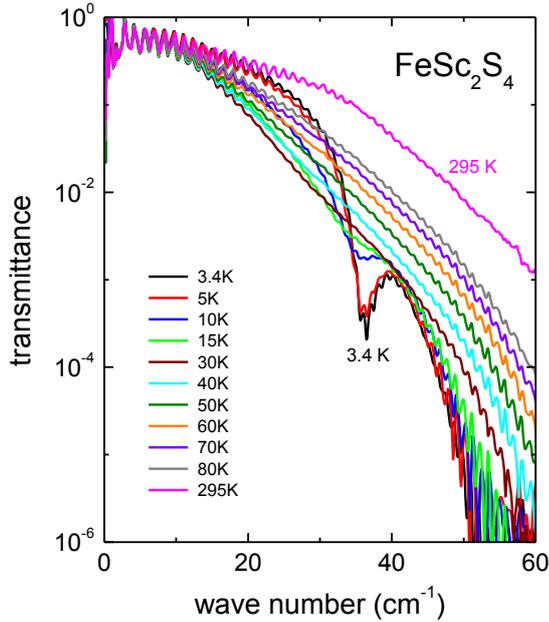

FIG. S1. THz Transmittance of FeSc$_2$S$_4$. Transmittance on a logarithmic scale as function of wave numbers up to 60 cm$^{-1}$ for room temperature, compared to a series of temperatures between 3.4 and 80 K in the spin-orbital liquid state.

Figure S1 shows the THz transmittance as measured for wave numbers between some cm$^{-1}$ up to 60 cm$^{-1}$ and temperatures 3.4 K < T < 80 K. The room-temperature transmittance is included for comparison. All spectra are modulated by interference patterns which become strongly enhanced at low and high wave numbers. They result from multiple reflections in the cryostat windows and from the sample and cannot be fully removed even by a detailed and advanced analysis including Fresnel optical formulas for reflectance and transmittance. Despite these experimental limitations, two remarkable facts can be clearly identified: i) At low temperatures an excitation evolves close to 30-35 cm$^{-1}$. Below about 15 K, it starts to shift to higher frequency and its damping is significantly reduced. For T > 50 K this excitation is either overdamped or has lost its dielectric oscillator weight. ii) While at 10 cm$^{-1}$ the transmittance remains almost temperature independent, at 50 cm$^{-1}$ the transmittance decreases by more than three orders of magnitude between room and liquid-helium temperature which makes it almost impossible to cover a large frequency regime by a single experimental set up. iii) It is also interesting to note how the transmittance close to 20 cm$^{-1}$ increases again below 80 K and almost reaches the room temperature value at 3.4 K. This enormous and non-continuous temperature dependence of the absorbance at this wave number range far below the phonon eigenfrequencies documents the strong coupling of spin-orbital excitations to electromagnetic radiation. The lowest-frequency phonon mode appears close to 115 cm$^{-1}$ and cannot explain the strong decrease in transmittance on decreasing temperatures.

For a closer analysis, the measured THz spectra, transmittance and phase shift, were converted into a wave-number dependence of the complex dielectric constant. In Fig. S2 we show the dielectric loss for selected temperatures between 4 and 70 K in a color-coded contour plot. These data have been taken on the same polycrystalline sample with the same thickness of 1.15 mm as those shown in Fig. S1, but were measured with different window configurations, to avoid interference noise close to the excitation frequencies, but also to prove reproducibility of the THz results, specifically the linear increase of dielectric loss corresponding to an excitation continuum of the spin orbital liquid. In Fig. S2 the evolution of a well-defined excitation close to 35 cm$^{-1}$ is clearly visible. The eigenfrequencies shift towards low wave numbers and reveal a strongly decreasing damping below about 15-20 K. The spin-orbital excitation appears on top of an almost linearly increasing background. The color code above Figure S2 indicates this linear background, with a dielectric loss which is zero at zero wave number and has been set to 0.8 at 65 cm$^{-1}$. The code pattern is very close to the frequency dependence of the loss at 70 K indicating that indeed at this temperature the loss increases strictly linearly. The additional enhancement of the loss at large wave numbers and low temperatures may provide experimental evidence of the appearance of a second mode outside of the covered frequency window.

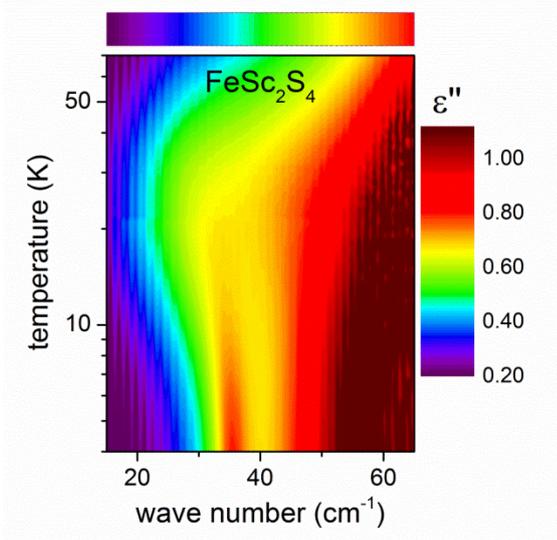

FIG. S2. Color coded contour plot of the dielectric loss of FeSc$_2$S$_4$. The dielectric loss is plotted vs. the logarithm of temperature (4 K < T < 70 K) and vs. wave number between 15 and 65 cm$^{-1}$ in a linear color code as indicated on the right side of Figure S2. The upper bar on top of the figure indicates a strictly linear increase of the loss with wave number.

## 2. FIR spectroscopy

As mentioned in the main article, the present reflectivity results could be satisfactorily described utilizing a fit routine developed by Kuzmenko, where Fano effects have been recently included[1]. To demonstrate the influence of the Fano parameter $\omega_q$ on the frequency dependence of the dielectric loss, Fig. S3 shows curves calculated with eq. (1) of the main paper for a series of Fano parameters. It is clear from eq. (1) that for $\omega_q = 0$ the Lorentzian line shape is recovered.

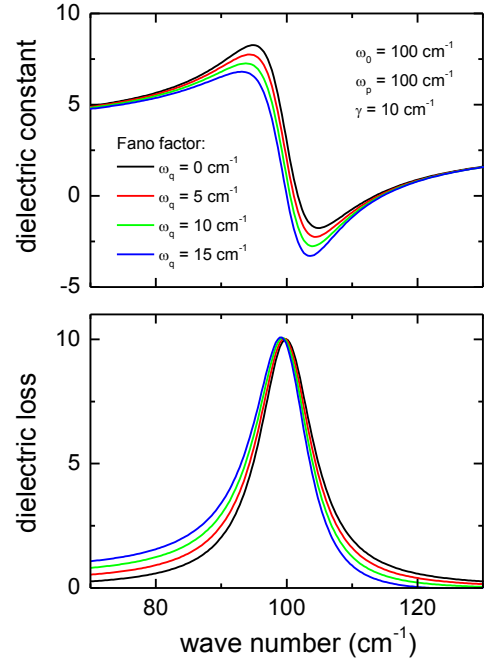

FIG. S3. Dielectric constant and loss calculated with different strengths of the Fano parameter. The complex dielectric constant is calculated for a model phonon as indicated in the upper frame using eq. (1) with different strengths of the Fano factor $\omega_q$.

As discussed in the main article, utilizing this fit routine the experimentally obtained reflectivity spectra for wavenumbers between 80 and 550 cm$^{-1}$ were fitted for all temperatures. Eigenfrequency $\omega_o$, damping $\gamma$, and oscillator strength $\Delta\varepsilon$ were used as free fit parameters for each phonon mode. The Fano factor $\omega_q$ was added to the Lorentzian oscillator and was treated as fit parameter for phonon mode 1. In addition, the high-frequency dielectric constant $\varepsilon_\infty$ was also treated as free parameter. As documented in Figs. 4(a) and 5 of the main paper, a reasonable agreement of fit and experimental data could be achieved in this way. Even at the lowest temperatures we found no indications of a splitting of phonon modes due to antiferromagnetic exchange, which is a universal feature for antiferromagnetic compounds[2]. From these fits we determined eigenfrequencies, dipolar strengths and damping constants for all four modes and, in addition for phonon 1, the Fano factor. The fit parameters eigenfrequency $\omega_0$, damping $\gamma$, oscillator strength $\Delta\varepsilon$, and Fano factor $\omega_q$ were determined for phonon mode 1 for all temperatures. The results are shown in the main manuscript in Figs. 4(b) - (e). For the modes 2 - 4, the temperature dependencies of eigenfrequeny, oscillator strength, and damping are documented in Fig. S4.

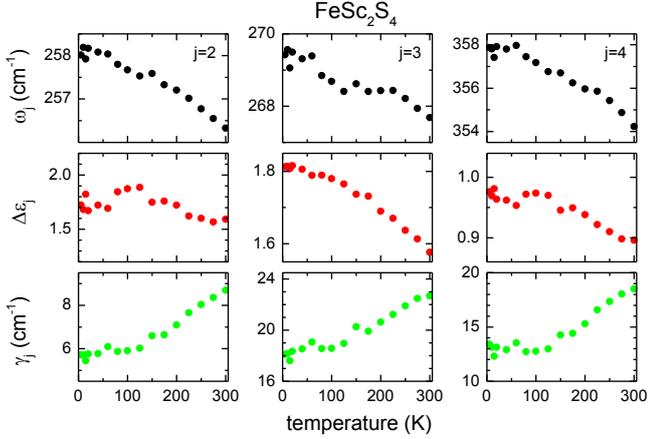

FIG. S4. Eigenfrequencies, oscillator strength, and damping of phonon modes 2 – 4 in $FeSc_2S_4$ as function of temperature.

The temperature dependence of eigenfrequencies and damping of these modes behave conventional, like in moderately anharmonic solids: The eigenfrequencies approach constant values at low temperatures (low compared to an overall Debye temperature) and decrease linearly on increasing temperatures, which mainly results from thermal expansion effects. The damping is low and constant at low temperatures and increases linearly on increasing temperatures. The oscillator strength is expected to be constant and temperature independent. We see continuous shifts of order 5%, which seem to be outside of experimental uncertainties and remain unexplained. They could only be explained assuming valence changes of the ions, which, however, seem rather unlikely in the strictly insulating material.

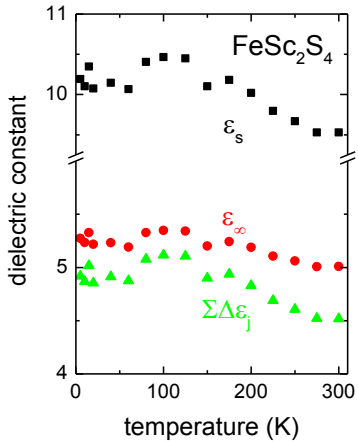

FIG. S5. Temperature dependence of static and high-frequency dielectric constants of $FeSc_2S_4$.

From the fits we also determined the temperature dependence of the high-frequency dielectric constant $\varepsilon_\infty$. As documented in Fig. S5, it is only weakly temperature dependent and of order 5. $\varepsilon_\infty$ and the sum over all four oscillator strengths $\Delta\varepsilon$ determine the temperature dependence of the static dielectric constants $\varepsilon_s$. As also documented in Fig. S5, the latter again is only weakly temperature dependent with values close to 10.

## 3. Experimental Details

**THz spectroscopy.** Time-domain THz transmission experiments were carried out on polycrystalline samples in the spectral range from 10 - 100 $cm^{-1}$ using a TPS Spectra 3000 spectrometer (TeraView Ltd.) on samples with thickness ranging from 0.7 to 1.2 mm and cross sections of the order 25 $mm^2$. Of course, the upper limit of the transmission spectra strongly depends on sample thickness and sample absorption, which in many cases strongly varies as function of temperature. Different cryostat windows of quartz glass with thickness of approximately 1 mm and of polypropylene with thickness of 75 μm were utilized. A He-flow cryostat was used to reach the temperatures ranging from 4 - 300 K. Transmission and phase shift were obtained from the Fourier-transformed time-domain signal. Dielectric constant and loss were calculated from transmission and phase shift by modelling the sample as a homogeneous dielectric. Here, for simplicity reasons, we assumed a magnetic permeability of $\mu = 1$, which may lead to some ambiguities in the case of purely magnetic contributions to the optical response.

**FIR experiments.** Reflectivity experiments were carried out in the FIR range using the Bruker Fourier-transform spectrometer IFS 113v equipped with a He-flow cryostat. With the set of mirrors and detectors used for these experiments, we were able to cover a frequency range from 60 to 700 $cm^{-1}$. For the reflectivity experiments, the ceramics were pressed with a maximum pressure of 1 GPa with the surfaces polished to optical quality. Nevertheless, neither surface nor density of the samples was ideal and of perfect optical quality and we were not able to receive reliable results below 80 $cm^{-1}$. In THz spectroscopy, real and imaginary part of the dielectric permittivity can be directly derived from transmission and phase shift. From the THz data, we determined absolute values of the real and imaginary parts of the permittivity. From the FIR reflectivity, the permittivity was calculated via a Kramers-Kronig transformation. We had to correct all FIR reflectivity spectra by a factor of 1.3 to bring in line the permittivity from the FIR and THz results. This correction was necessary to account for low densities and non-ideal surfaces of the ceramic samples. In addition, we produced high-density samples, with large grain size and almost perfect surface quality by spark-plasma sintering. In these samples, we found good agreement with the absolute values of the permittivity as determined by THz spectroscopy. The obtained permittivity was approximately by a factor of 1.3 larger than for the unscaled FIR results of standard ceramic samples, justifying the applied factor. However, in these samples dc-conductivity contributions dominated at low frequencies revealed by THz spectroscopy, probably due to slight off-stoichiometries of sulfur and, hence, considerable doping effects. For the combined data, it seems necessary to use results obtained on the same samples and, hence, in the FIR region the scaled data set is used. All reflectivity spectra were fitted using the RefFIT fit routine[1], including the option for Fano-type resonance line shapes.